# NbTiN superconducting nanowire detectors for visible and telecom wavelengths single photon counting on Si3N4 photonic circuits

*C. Schuck, W. H. P. Pernice[*], and H. X. Tang*

*Department of Electrical Engineering, Yale University, New Haven, CT 06511, USA*
[*]*Current address: Institute of Nanotechnology, Karlsruhe Institute of Technology, 76133 Karlsruhe, Germany*

We demonstrate niobium titanium nitride superconducting nanowires patterned on stoichiometric silicon nitride waveguides for detecting visible and infrared photons. The use of silicon nitride on insulator on silicon substrates allows us to simultaneously realize photonic circuits for visible and infrared light and integrate them with nanowire detectors directly on-chip. By implementing a traveling wave detector geometry in this material platform, we achieve efficient single photon detection for both wavelength regimes. Our detectors are an ideal match for integrated quantum optics as they provide crucial functionality on a wideband transparent waveguide material.

Interfacing optical circuitry and high-efficiency single photon detectors with low loss is one of the key challenges of quantum photonic technologies [1,2]. Ideally, these components are integrated with non-classical light sources on a scalable monolithic platform [3]. To satisfy the source and circuit demands makes it highly desirable to work with a material system which allows for simultaneous operation at visible and infrared wavelengths. This is achievable by using silicon nitride (SiN) waveguides with low optical absorption allowing for low-loss wave-guiding both in the infrared [4] and visible [5] wavelength regime. The recent discovery of a surface $\chi^{(2)}$ effect in silicon nitride [6] also brings the possibility of creating non-classical on-chip photon sources into reach. On the other hand, superconducting nanowire single-photon detectors (SSPD) are well suited for the integration with nanophotonic circuitry and offer superior performance compared to more traditional detector technologies [7,8]. The detection mechanism is based on the superconducting to normal state transition of a nanowire segment, induced by the absorption of a single photon creating a localized hot-spot [9]. Most state-of-the-art SSPDs are implemented as niobium nitride (NbN) meander wires designed as stand-alone units coupled to optical fibers and achieve high speed operation with high timing accuracy and sensitivity over a broad spectral range [9,10,11]. These detectors are optimized for high system efficiency as desired for free-space or optical fiber environments, e.g. in quantum key distribution [12] or high-data-rate optical (space) communication [13,14]. The waveguide coupled SSPDs presented here is a development primarily addressing the needs of integrated photonics and quantum information processing [7,15].

We demonstrate a material system based on niobium titanium nitride (NbTiN) SSPDs on SiN waveguides to achieve high on-chip detection efficiency for visible and infrared photons on a scalable platform. NbTiN thin films are attractive for such an implementation because they can be deposited with high homogeneity and are more versatile in growth conditions as compared to other superconducting materials [10,17,18]. This also allows for achieving lower dark count rates with NbTiN SSPDs as compared to NbN nanowire detectors at similar detection efficiency [10,17,19]. Furthermore, NbTiN may be a preferable material choice over NbN for achieving higher detector speed [18,20]. The highly efficient detection of single photons achieved here is a result of patterning the detectors directly on top of the SiN waveguides in a traveling wave design [16]. This allows us to realize very large interaction lengths of the guided light field with the nanowire on the waveguide surface. For a sufficiently long nanowire the incoming photons are fully absorbed in contrast to traditional fiber-coupled meander-type SSPDs which absorb photons in a few nanometer thin film under normal incidence [11,17].

We fabricate two types of SiN photonic circuits on the same chip, one for operation at 775nm and one for 1550nm wavelength, as shown in Fig. 1 (a). Each circuit is composed of a pair of optical grating couplers, a 50:50 waveguide

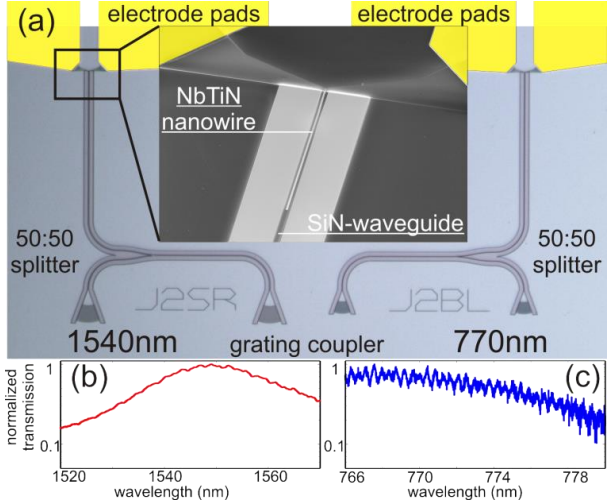

**Figure 1** (a) Silicon nitride photonic circuits consisting of a calibration couplers, 50:50 splitters and low loss waveguides are fabricated on the same chip for operation in the telecom band (left) and visible (right) wavelength regime. The inset shows the detector region for the telecom band photonic circuit, where a U-shaped NbTiN nanowire is patterned on top of the SiN-waveguide. The nanowire ends are connected via NbTiN patches to a pair of electrode pads for electrical access. (b) infrared light spectrum obtained by sending light from a tunable laser source into the grating coupler input port (right on infrared device) and recording the power transmitted at the reference output port (left on infrared device) (b) visible light spectrum obtained by sending light from a tunable laser source into the grating coupler input port (left on visible device) and recording the power transmitted at the reference output port (right on visible device).

splitter and the detector region which connects to a pair of electrode pads. The grating couplers are used to couple light from an optical fiber array into single mode waveguides on the chip. Their operating wavelength (visible vs. infrared) is adjusted by optimizing the period and filling factor of the grating. The 50:50 waveguide splitter [21] routes half of the coupled light to the detector region and the other half to a reference port which allows us to precisely determine the number of photons arriving at the detector region. Transmission spectra recorded at the calibration port are shown in Fig 1 (b) for the telecom band and in Fig. 1 (c) for the visible light devices, showing broadband transmission in both wavelength regimes. After careful alignment of the chip with respect to the fiber array using low-temperature compatible nano-positioners we observe -10.9 dB transmission loss per coupler on the infrared device measured at 1542nm and -13.1 dB loss per coupler on the visible device determined at 768nm input wavelength.

The devices are realized by patterning stoichiometric silicon nitride on silicon wafers covered with a buried oxide layer (BOX). The SiN layer has a thickness of 330nm to achieve single (TE) mode waveguide operation in both wavelength regimes by adjusting the waveguide widths. In combination with the underlying 3.3 μm BOX layer, which acts as a cladding, we thus achieve low optical transmission loss and optimal coupling efficiency for the on-chip input/output grating couplers. The detectors are fabricated from an 8nm thin NbTiN film which is deposited by dc magnetron sputtering on top of the SiN layer. The sheet resistance of the as deposited 8nm film was measured as 231 Ω/sq. Electrode pads and markers for alignment of the subsequent detector and waveguide patterns are defined in a first electron beam lithography step using polymethyl methacrylate (PMMA) positive resist. We then employ electron-beam evaporation of a 6nm Cr adhesion-layer and a 200nm Au-film followed by lift-off in acetone to realize the contact pads and alignment markers. High resolution electron beam lithography on hydrogen silsesquioxane (HSQ) resist defines the nanowire detector features which are subsequently transferred into the NbTiN layer in a timed reactive ion etch (RIE) step using CF4 chemistry. The waveguide layer is then patterned in a final electron beam lithography step employing ZEP520A positive resist and a carefully timed RIE through the SiN to the interface with the buried oxide layer using CHF3 chemistry. The inset of Fig. 1 (a) shows a scanning electron microscope (SEM) image of the detector region after removal of the ZEP resist. The U-shaped nanowire is protected by a thin layer of HSQ resist which remained after the NbTiN etch, as visible on top of the waveguide. From the SEM images we estimate the alignment tolerance of the electron beam pattern generator (Vistec EBPG 5000+ 100kV) to be better than 50nm which allows for centering the nanowire on the waveguides with high accuracy. The nanowire-ends are connected via large NbTiN patches to ground and signal electrode pads. Electrical readout of the SSPDs is realized by approaching an RF-probe to the contact pads.

A schematic of the traveling wave design is shown in Fig. 2 (a). The NbTiN nanowire on top of the waveguide has a thickness of 8nm, a wire width of 75nm, a wire spacing of 90nm and length of 40μm (footprint). The SiN waveguides are designed to support only the fundamental quasi-TE mode at 775nm and 1550nm for visible and infrared devices, respectively. Figure 2 (b) and 2 (f) show calculations of the electrical component of the optical field distribution for the fundamental mode profiles of 775nm light in a 330x600nm$^2$ SiN waveguide and 1550nm light in a waveguide of 330x1000nm$^2$ cross section, respectively. The strong spatial mode confinement due to the high refractive index contrast between the waveguide material (SiN) and its environment (vacuum / oxide) allows for very compact optical device design as desired for large scale photonic circuit implementations. Similar simulations for the waveguide

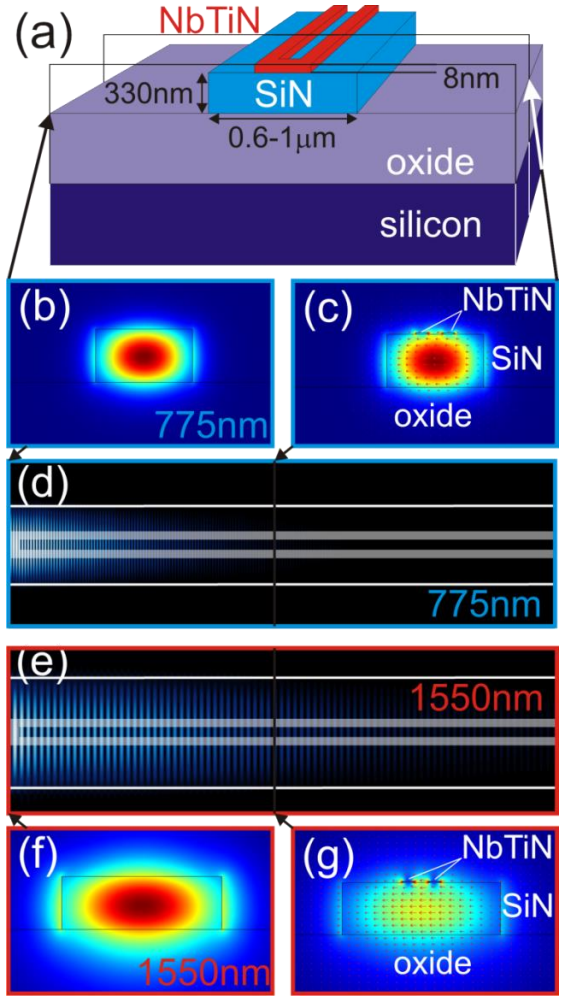

**Figure 2** (a) Schematic of the traveling wave detector design. The SiN waveguide has a width of 600nm (1μm) in the visible (infrared) light design. The NbTiN wire of 8nm height and 75nm wire width is patterned directly on top of the waveguide to sense the evanescent light field over a length of 40μm; (b) simulated electrical field distribution (normalized) of the optical mode (TE, Ex) for 775nm wavelength in the bare 600nm wide SiN waveguide; (c) simulated electrical field distribution (normalized) of the optical mode (TE, Ex) for 775nm wavelength in the NbTiN nanowire covered region of the 600nm wide SiN waveguide; (d) FDTD simulation of the absorption in the NbTiN nanowire covered waveguide region (600nm width) for 775nm; (e) FDTD simulation of the absorption in the NbTiN nanowire covered waveguide region (1μm) for 1550nm; (f) simulated electrical field distribution (normalized) of the optical mode (TE, Ex) for 1550nm wavelength in the bare 1μm wide SiN waveguide; g) simulated electrical field distribution (normalized) of the optical mode (TE, Ex) for 1550nm wavelength in the NbTiN nanowire covered region of the 1μm wide SiN waveguide. Light intensities are shown in linear color scale.

sections which carry the superconducting thin film show how the evanescent tail of the light field couples strongly to the NbTiN nanowire. In Fig. 2 (c) and 2 (g) we observe field enhancement at the nanowire sides both for the 775nm and 1550nm field distributions, respectively. The large change in the imaginary part of the refractive index from the bare waveguide to the NbTiN covered region then leads to strong absorption of the incoming light. The real part of the refractive index on the other hand only changes by 0.7% (0.1%) for the 1550nm (775nm) mode such that the transition from the bare to the nanowire covered waveguide section is smooth. Negligible reflection at this interface is also expected from the large difference in SiN to NbTiN film cross section (103:1 for the 600nm wide waveguide and 172:1 for the 1μm wide waveguide) leading only to a small spatial expansion of the mode profile in the NbTiN covered waveguide section.

A numerical analysis of the NbTiN nanowire absorption properties is presented in Fig. 2 (d) for 775nm and in Fig. 2 (e) for 1550nm light. The three-dimensional finite-difference time-domain (FDTD) simulations yield an absorption rate of 0.33 dB/μm for 1550nm light in the NbTiN nanowire covered region of the waveguide, and 0.71 dB/μm for the 775nm case. The stronger absorption value for visible light is also consistent with the larger imaginary part obtained from the calculations of the visible light mode in the NbTiN covered waveguide section as compared to the infrared mode (Fig. 2 (c) and (g)). These absorption rates are lower than those obtained for NbN nanowires on silicon waveguides [7] due to the larger cross section of the SiN waveguides used here. We estimate an absorption of 95.1% (99.8%) of the incident 1550nm (775nm) photons along our 40μm long SSPD. Such high absorption efficiency (AE) is a prerequisite for high on-chip detection efficiency OCDE = AE * QE, where QE is the quantum efficiency of the detector.

The fabricated devices are then characterized experimentally. The optical measurement setup consists of continuous wave lasers for 1520-1570 nm (New Focus 6428) and 765-780 nm (New Focus TLB 6712) wavelength launching light via calibrated optical attenuators (for 1550 nm / 775 nm), polarization controllers and a fiber array into a liquid helium (LHe4) cryostat fitted with a 1K refrigerator. The sample is mounted on a stack of nano-positioners inside the cryostat such that the grating couplers (see Fig. 1) can be aligned to the fibers for optimal transmission through the device. The optical output from the calibration coupler port is collected with another optical fiber in the same fiber array and detected with a calibrated photo-detector. The nanowire detector is current biased and read out via an RF-probe which is brought into contact with the electrode pads on the chip (Fig. 1). Using a programmable current source (Keithley 2400) connected to the RF-probe via a 10 kHz low-pass filter and a 6 GHz bias-T we scan the bias current across the nanowire at a temperature of 1.57 K and observe switching currents of 17.76 μA for both detectors on the 775 nm and 1550 nm devices. The close match of the critical current values for two independent nanowire detectors highlights the high reproducibility of our NbTiN nanowire fabrication process across the chip.

To calculate the on-chip detection efficiency it is necessary to determine the number of photons inside the waveguide reaching the SSPD, which is given by $N_d = P_{in} \times T_{gc,in} \times T_{50:50} \times T_{ext}/E_\gamma(\lambda)$. Here the attenuation by the input grating coupler $T_{gc,in}$ can be determined by measuring the transmission at the reference port $P_{ref} = P_{in} \times T_{gc,in} \times T_{50:50} \times T_{gc,ref}$ for input power $P_{in}$. Due to the optical reciprocity of the optical grating couplers the transmission loss at the input, $T_{gc,in}$, is identical to that at the reference port, $T_{gc,ref}$, as confirmed on separate calibration devices. Accounting furthermore for transmission through the 50:50 waveguide splitter, $T_{50:50}$, we hence obtain the on-chip optical power from the measured laser input power, $P_{in}$, after adding external attenuation on the order of $T_{ext} = 90 - 100\ dB$. The exact attenuation value is determined separately for each measurement series both for 768 nm and 1542 nm input light by using the same calibrated photo-detector as above. Note that the linearity of the attenuators from typical laser input power to operation at single-photon level has been confirmed independently [7]. The number photons arriving at the detector then follows from dividing the on-chip optical power with the photon energy $E_\gamma(\lambda = 768nm) = 2.587 \times 10^{-19} J$ and $E_\gamma(\lambda = 1542nm) = 1.288 \times 10^{-19} J$, respectively. The dark count rate for both detectors was below 10 Hz and hence does not contribute appreciable to the number of photons arriving at the detector (approximately five orders of magnitude higher). We find the dark counts to be caused mainly by straylight coupled through the cladding of the fibers leading into the cryostat when the laser is disconnected, as also confirmed by variations of the dark count rate with the ambient light level.

The detector output pulses are amplified with low-noise high-bandwidth amplifiers and recorded on a single photon counting (SPC) system (PicoHarp 300). The high signal-to-noise ratio of the output pulses after amplification allows us to clearly distinguish photon detection events from electrical noise. Using a high-bandwidth oscilloscope we find the output-pulse width as 1.4ns at full width half maximum for these detectors of 2x 40 μm length by using low-noise amplifiers with small low frequency cut-off values which cause only minor output voltage overshoot. Operation in the single photon regime is further confirmed by observing a linear dependence of the photon count rate on the photon flux directed to the device. We then obtain the on-chip photon detection efficiency as a function of bias current from the ratio of recorded detector output pulses to the on-chip photon number (see above) as shown in Fig. 3. For infrared photons of 1542 nm wavelength we find a maximal on-chip detection efficiency of 52.5% measured at 99% of

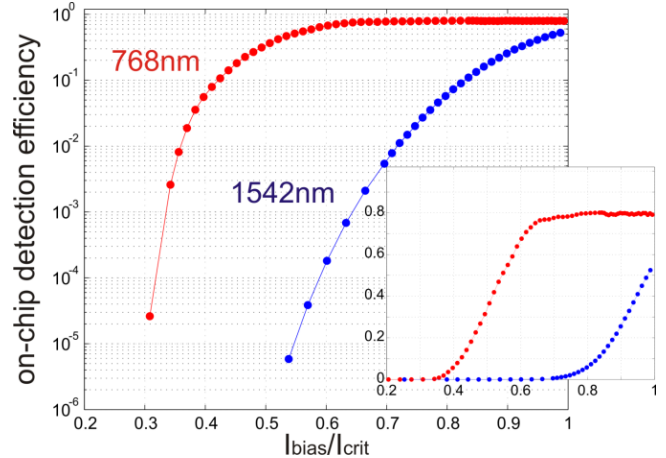

**Figure 3** (color online) On-chip detection efficiency for NbTiN nanowire SSPDs with 75nm wire width and 40 μm length (footprint) as a function of bias current in units of the critical current for 768nm photons (red) and for 1542nm photons (blue). Inset: The same data on a linear scale reveals how the on-chip detection efficiency for 768nm photons saturates around 80% over a bias current range of 70-100% of the critical current, while it monotonically increases to 52.5% at 99% Icrit for 1542nm photons.

the critical current. Similarly, for visible light (768nm) the corresponding detector achieves a maximal on-chip detection efficiency of 80.1%. In this case we observe how the on-chip detection efficiency saturates around 80% starting from a bias current of 70% of the critical current. The inset of Fig. 3 shows the on-chip detection efficiency as a function of bias current on linear scale highlighting the saturation behavior for the visible photon detection rate whereas the detector does not reach saturation for infrared photons. The saturation behavior is explained by the linear dependence of the hotspot diameter on the incident photon energy [22]. At high bias current a 768 nm photon provides sufficient energy to always causes a hotspot which is large compared to the wire dimensions such that every photon absorbed in the superconducting film leads to an output pulse of the SSPD. The detection of lower energy 1542 nm photons on the other hand does not reach high quantum yield in the nanowire used here, even when biased very close to the switching current, as expected from the exponential dependence of quantum efficiency on wavelength [23,24]. A slightly higher absolute detector sensitivity for visible light in our waveguide coupled SSPDs is furthermore expected according to the above reported absorption characteristics for 780nm and 1550nm light (see Fig. 2). We anticipate that detector performance can be optimized by engineering the nanowire dimensions.

The uncertainty of the measured detection efficiency values is determined by the uncertainty of the 50:50 waveguide splitting and the optical power measurements employed in the grating coupler and attenuator calibration procedures. We obtain a 5.9% total uncertainty of the photon number arriving at the detector after propagating the 4% splitter uncertainty, the 3% calibration uncertainty of our photo-detector (Newport 818-IS-1, NIST traceable) and a 0.5% linearity uncertainty through the grating coupler and attenuator calibration procedure. Here we do not account for changes of the on-chip photon number due to polarization drift. Since we optimize the device transmission at the reference port before each measurement run (including calibration procedures) the on-chip detection efficiency values given above are worst case numbers in this respect. We note however that no significant polarization drifts are observed once the cryostat has reached stable operation temperature. This is also evident from the data presented in the inset of Fig. 3 which does not show significant changes in detector count rate in the saturated regime over the course of the measurement time of 10-1 seconds per data point.

In conclusion, we have integrated superconducting single photon detectors with photonic circuits for visible and infrared light on a scalable platform for chip-scale quantum optical technology. We demonstrate high on-chip detection efficiency of 52.5% for telecom band photons and 80.1% for visible wavelength photons over a large bias current range, achieved in a NbTiN on SiN material system. The efficient interface between high-performance single photon detectors, visible and infrared photonic waveguides constitutes a crucial ingredient for developing future large scale quantum optical implementations.

We thank Robin Cantor for preparing the NbTiN thin films. W.H.P. Pernice acknowledges support by the DFG grant PE 1832/1-1. H.X.T acknowledges support from a Packard Fellowship in Science and Engineering and a CAREER award from the National Science Foundation. We want to thank Dr. Michael Rooks and Michael Power for their assistance in device fabrication.

1. J. L. O'Brien, Science **318**, 1567 (2007).
2. C. M. Natarajan, A. Peruzzo, S. Miki, M. Sasaki, Z. Wang, B. Baek, S. Nam, R. H. Hadfield, J. L. O'Brien, Appl. Phys. Lett. **96**, 211101 (2010).
3. A. Politi, J. C. F. Matthews, M. G. Thompson, J. L. O'Brien, IEEE J. Sel. Top. Quantum Electron. **15**, 1673 (2009).
4. M.-C. Tien, J. F. Bauters, M. J. R. Heck, D. T. Spencer, D. J. Blumenthal, J. E. Bowers, Opt. Express 19, 13551 (2011).
5. E. Shah Hosseini, S. Yegnanarayanan, A. H. Atabaki, M. Soltani, A. Adibi, Opt. Express **17**, 14543 (2009).
6. J. S. Levy, M. A. Foster, A. L. Gaeta, M. Lipson, Opt. Express **19**, 11415 (2011).
7. W. H. P. Pernice, C. Schuck, O. Minaeva, M. Li, G. N. Goltsman, A. V. Sergienko, H. X. Tang, Nat. Comm. arxiv:1108.5299
8. R. H. Hadfield, Nat. Photon. **3**, 696 (2009).
9. G. N. Gol'tsman, O. Okunev, G. Chulkova, A. Lipatov, A. Semenov, K. Smirnov, B. Voronov, A. Dzardanov, C. Williams, R. Sobolewski, Appl. Phys. Lett. **79**, 705 (2001).
10. S. N. Dorenbos, E. M. Reiger, U. Perinetti, V. Zwiller, T. Zijstra, T. M. Klapwijk, Appl. Phys. Lett. **93**, 131101 (2008).
11. A. Gaggero, S. Jahanmiri Nejad, F. Marsili, F. Mattioli, R. Leoni, D. Bitauld, D. Sahin, G. J. Hamhuis, R. Nötzel, R. Sanjines, A. Fiore Appl. Phys. Lett. **97**, 151108 (2010).
12. H. Takesue, S. W. Nam, Q. Zhang, R. H. Hadfield, T. Honjo, K. Tamaki, Y. Yamamoto, Nat. Photon. **1**, 343 (2007).
13. D. M. Boroson, J. J. Scozzafava, D. V. Murphy, B. S. Robinson, H. Shaw, Third IEEE Intl. Conf. on Space Mission Challenges for Info. Technol., pp 23-28, (2009).
14. B. S. Robinson, A. J. Kerman, E. A. Dauler, R. J. Barron, D. O. Caplan, M. L. Stevens, J. J. Carney, S. A. Hamilton, J. K. Yang, K. K. Berggren, Opt. Lett. **31**, 444 (2006).
15. J. P. Sprengers, A. Gaggero, D. Sahin, S. Jahanmirinejad, G. Frucci, F. Mattioli, R. Leoni, J. Beetz, M. Lermer, M. Kamp, S. Höfling, R. Sanjines, A. Fiore, Appl. Phys. Lett. **99**, 181110 (2011).
16. X. Hu, C. W. Holzwarth, D. Masciarelli, E. A. Dauler, K. K. Berggren, IEEE Trans. Appl. Supercond. **19**, 336 (2009).
17. M. G. Tanner, C. M. Natarajan, V. K. Pottapenjara, J. A. O'Connor, R. J. Warburton, R. H. Hadfield, B. Baek, S. Nam, S. N. Dorenbos, E. Bermúdez Ureña, T. Zijstra, T. M. Klapwijk, V. Zwiller, Appl. Phys. Lett. **96**, 221109 (2010).
18. S. Miki, M. Takeda, M. Fujiwara, M. Sasaki, A. Otomo, Z. Wang, Appl. Phys. Express **2**, 075002 (2009).
19. M. K. Akhlaghi, H. Atikian, A. Eftekharian, M. Loncar, A. H. Majedi, Opt. Express **20**, 23610 (2012).
20. A. J. Annunziata, O. Quaranta, D. F. Santavicca, A. Casaburi, L. Frunzio, M. Ejrnaes, M. J. Rooks, R. Cristiano, S. Pagano, A. Frydman, D. E. Prober, J. Appl. Phys. **108**, 084507 (2010).
21. K. Y. Fong, W. H. P. Pernice, M. Li, H. X. Tang, Appl. Phys. Lett. 97, 073112 (2010).22.
22. A. Verevkin, J. Zhang, R. Sobolewski, A. Lipatov, O. Okunev, G. Chulkova, A. Korneev, K. Smirnov, G. N. Gol'tsman, A. Semenov, Appl. Phys. Lett. 80, (2002).
23. A. Korneev, V. Matvienko, O. Minaeva, I. Milostnaya, I. Rubtsova, G. Chulkova, K. Smirnov, V. Voronov, G. Gol'tsman, W. Słysz, A. Pearlman, A. Verevkin, R. Sobolewski, IEEE Trans. Appl. Supercond. **15**, 571 (2005).
24. A. Semenov, A. Engel, K. Il'in, G. Gol'tsman, M. Siegel, H.-W. Hübers, Eur. Phys. J. AP 21, 171 (2003).